\documentclass[9pt,twocolumn]{extarticle}
\pdfoutput=1

\usepackage[T1]{fontenc}
\usepackage{mathpazo}   
\normalfont
\usepackage{courier}

\usepackage{amsmath,amssymb}

\usepackage[utf8]{inputenc}

\usepackage{graphicx}
\usepackage{subcaption}
\usepackage{capt-of}
\usepackage{caption}
\usepackage[margin=1in]{geometry}
\usepackage[usenames,dvipsnames]{color}

\usepackage[hyphens]{url}
\usepackage{hyperref}
\hypersetup{
  colorlinks=true,
  linkcolor=Blue,
  urlcolor=Blue,
  citecolor=Blue
}

\usepackage[
  backend=biber,
  style=nature,
  sorting=none,
  autocite=superscript, 
  maxbibnames=10,
  minbibnames=3,
  articletitle=true,
  doi=true, url=false,
  eprint=false, isbn=false,
  giveninits=true,
  date=year,
  defernumbers=true
]{biblatex}
\let\cite\supercite

\defbibheading{subbibliography}{\section*{#1}}

\DeclareFieldFormat{doi}{%
  \ifhyperref{\href{https://doi.org/#1}{https://doi.org/#1}}{https://doi.org/#1}}
\AtEveryBibitem{\iffieldundef{doi}{}{\clearfield{url}}}

\makeatletter
\renewbibmacro*{volume+number+eid}{%
  \printfield{volume}%
  \iffieldundef{number}{}{\printtext[parens]{\printfield{number}}}%
  \setunit{\addcomma\space}%
  \printfield{eid}}
\renewbibmacro*{journal+issuetitle}{%
  \printfield[journaltitle]{journaltitle}%
  \setunit*{\addspace}%
  \printfield{series}%
  \setunit*{\addspace}%
  \usebibmacro{volume+number+eid}%
  \iffieldundef{volume}
    {\iffieldundef{number}{}{\printtext[parens]{\printfield{number}}}}
    {}%
  \newunit}
\makeatother
\usepackage{soul}
\usepackage{titlesec}

\titleformat{\section}[block]
  {\bfseries\sffamily\normalsize}   
  {\thesection}{0.6em}{}            
\titlespacing*{\section}{0pt}{0.6\baselineskip}{0.35\baselineskip}

\titleformat{\subsection}[block]
  {\bfseries\sffamily\small}
  {\thesubsection}{0.6em}{}
\titlespacing*{\subsection}{0pt}{0.5\baselineskip}{0.25\baselineskip}

\titleformat{name=\section,numberless}[block]
  {\bfseries\sffamily\normalsize}{}{0pt}{}

\usepackage{dsfont}
\usepackage{upgreek}
\usepackage[squaren]{SIunits} 
\usepackage[normalem]{ulem}  

\makeatletter\def\blx@maxline{77}\makeatother

\def\mytitle{Long-range propagating paramagnon-polaritons\\ in organic free radicals} 
\title{\vspace{-1.5cm}\Huge\textbf{\textrm{\mytitle}}} 

\author{Sebastian~Knauer$^{1*}$, Roman~Verba$^{2}$, Rostyslav~O.~Serha$^{1,3}$, Denys~Slobodianiuk$^{2}$,\\   David Schmoll$^{1,3}$, Andreas Ney$^{4}$, Sergej~O.~Demokritov$^{5}$, Andrii~V.~Chumak$^{1*}$}

\date{} 
\begin{document}

\twocolumn[{
\maketitle 
\vspace{-8mm}
\begin{center}
\begin{minipage}{0.95\textwidth}
\begin{center}
\textit{\textrm{
\textsuperscript{1} Faculty of Physics, University of Vienna, 1090 Vienna, Austria.\\
\textsuperscript{2} V. G. Baryakhtar Institute of Magnetism of the NAS of Ukraine, 03142 Kyiv, Ukraine. \\
\textsuperscript{3} Vienna Doctoral School in Physics, University of Vienna, 1090 Vienna, Austria.\\
\textsuperscript{4} Institut für Halbleiter-und Festk\"orperphysik, Johannes Kepler Universit\"at, 4040, Linz, Austria.\\
\textsuperscript{5} Institute for Applied Physics, University of M{\"u}nster, 48149 M{\"u}nster, Germany.\\   
}}
\end{center}
\end{minipage}
\end{center}
}]

\noindent

\textbf{Keywords:} Paramagnonics, Propagating Spin-Wave Spectroscopy, Millikelvin Temperature Magnonics
\vspace{0.2cm}

\noindent

\begin{refsegment}
\noindent

\textbf{
Materials are commonly distinguished by their magnetic response into diamagnetic, paramagnetic, and magnetically ordered (ferro-, ferri-, and antiferromagnetic) phases. 
Diamagnets and paramagnets lack spontaneous long-range order, whereas ordered magnets develop such order below their Curie or N\'eel temperature and support single spin-wave excitations (magnons).
Magnons have found applications in radio-frequency technologies and computation~\cite{Cornelissen2015,Giribaldi2024,Zenbaa2025a,Zenbaa2025b}, magneto-optics~\cite{Pintus2022,Pintus2025}, and foundational quantum experiments~\cite{Demokritov2006, Morris2017, Lachance-Quirion2020, Schneider2021, Chumak2022, Dobrovolskiy2025}.
Above the Curie/N\'eel temperature, long-range order is lost and the material transitions to a paramagnetic phase, with localised spin alignment in small patches, producing paramagnons with only short-range propagation.
Here we show that long-range coherence is preserved in the organic free radical 2,2,6,6-tetramethylpiperidin-1-oxyl~\cite{Bordeaux1973} above the N\'eel temperature using all-electrical propagating spin-wave spectroscopy in external magnetic fields. 
We observe coherently excited low-energy paramagnon-polaritons up to \(\mathbf{23\,\mathrm{GHz}}\) , propagating over \(\mathbf{8\,\mathrm{mm}}\) at supersonic group velocities exceeding \( \mathbf{100\,\mathrm{km\,s^{-1}}}\). 
Using free radicals as magnon carriers integrates organic materials with spintronics and opens the way to organic electronics, dense information storage, and quantum technologies.
}


Early experiments detected non-propagating paramagnetic spin waves~(paramagnons) by inelastic neutron scattering, e.g., in CrMn~\cite{Sinha1969}, Pd~\cite{Doubble2010}, TlCuCl~\cite{Sushkov2014, Merchant2014}, MnTe~\cite{Zheng2019}, and in high-$T_{\mathrm{c}}$ cuprates such as YBaCuO~\cite{Wang2022}. These paramagnons typically carry $\geq$~10\,meV energy (frequency $\geq$~2\,THz) set by proximity to ordering and broad electronic linewidths.
Other studies probed spin-current transport in paramagnets via spin pumping~\cite{Shiomi2014, Okamoto2016}, the paramagnetic and triplon spin-Seebeck effects~\cite{Wu2015, Hirobe2017, Chen2021}, and direct/inverse spin Hall signals~\cite{Oyanagi2019} between Pt stripes on GGG. These observed diffusive signals demonstrate that paramagnets can carry spin information, but they lack the phase, amplitude–phase quadrature, and frequency–wavevector access necessary to establish coherent dispersion.

\begin{figure*}[t!]
\centering
\includegraphics[width=1\textwidth]{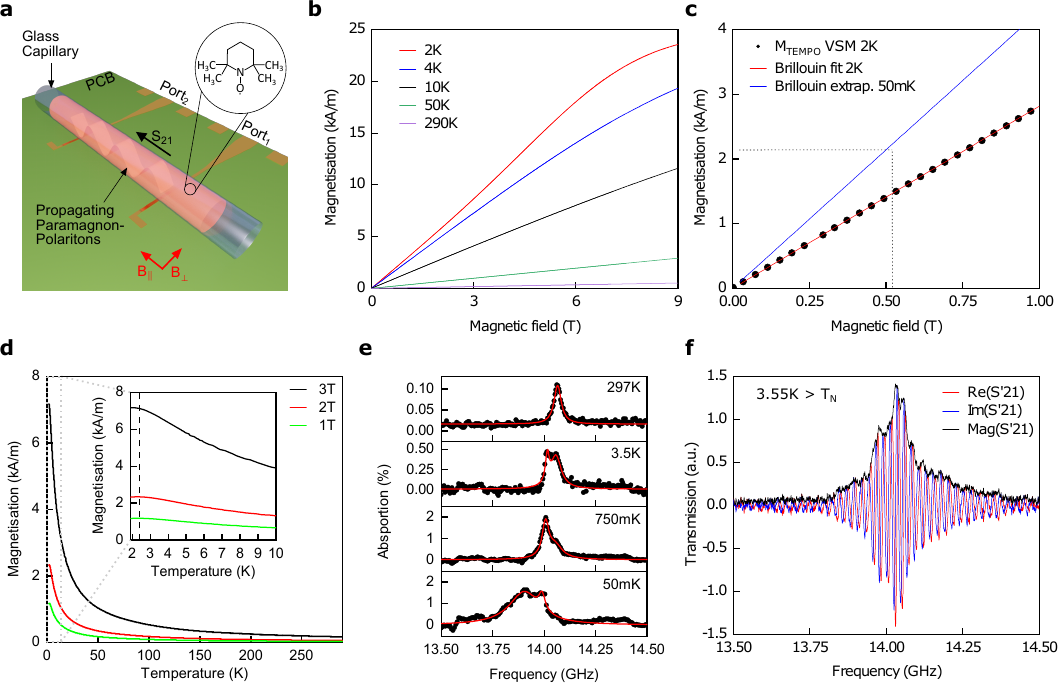}
\caption{~\textbf{TEMPO characterisation and propagation demonstration:}
\textbf{a,}~Sample schematic for all-electrical propagating spin-wave spectroscopy of paramagnon-polaritons. The microwave signal is applied on port~1 (antenna~1), exciting paramagnon-polaritons in a TEMPO-filled glass capillary and detected on port~2 (antenna~2), under an external field (B$_{\perp}$ or B$_{\parallel}$). Inset:~Chemical diagram of the aminoxyl radical C$_9$H$_{18}$NO~(TEMPO). It is characterised by its NO-group, with one unpaired electron delocalised over the nitrogen-oxygen bond.
~\textbf{b,}~Magnetisation of TEMPO with respect to the magnetic field for different temperatures, using vibrating sample magnetometry. The signal saturates for high magnetic fields and low temperatures at about 24\,$\mathrm{kA/m}$.~\textbf{c,}~Brillouin fit of the 2\,K measurement (in \textbf{b,}) and extrapolation for 50\,mK. For example, at 500\,mT an effective magnetisation of around 2.2\,$\mathrm{kA/m}$ can be estimated (dashed line).
~\textbf{d,}~Magnetisation with respect to temperature. The curves flatten below about 3\,K (inset), indicating the phase transition between paramagnetic and antiferromagnetic states. The black dashed line indicates the $T_{\mathrm{N}}=2.4\,\mathrm{K}$ literature value.
~\textbf{e,}~Electron paramagnetic resonance (EPR) absorption measurements of TEMPO comparing room, kelvin and millikelvin temperatures, taken at 500\,mT. The linewidth increases from 42\,MHz at 297\,K to about 67\,MHz at 750\,mK, while the frequency shifts from 14.066\,GHz to 14.014\,GHz (fit errors negligible). At lower temperatures, the linewidth broadens. A single Lorentz fit is used at room temperature, and a double Lorentz fit is used at cryogenic temperatures (see main text for more details). 
~\textbf{f,}~Demonstration of all-electrical propagating spin-wave spectroscopy transmission signal (S$'_{21}$) in the paramangetic state of TEMPO at 3.55\,K and at B$_{\parallel}=$~500\,mT (paramagnon-polariton). The linear magnitude~(black), real~(red), and imaginary part~(blue) of the transmission signal are shown. The phase of the propagation is preserved, as indicated by the $\pi/2$-phase relation between the real and imaginary parts.
}
\label{fig1}
\end{figure*}

Here we demonstrate that long-range magnonic order can be sustained in the stable organic free radical TEMPO (C$_9$H$_{18}$NO, 2,2,6,6-tetramethylpiperidin-1-oxyl;~Tanane/Tanan).
TEMPO is known for its use as a catalyst and chemical oxidant in producing pharmaceuticals, flavours, fragrances, and agrochemicals, and in non-linear optical materials and cathode materials for rechargeable lithium batteries~\cite{Takahashi1997, Nakahara2002, Ciriminna2010, Hoover2011}. It is characterised through its NO-group with one unpaired electron delocalised over the nitrogen-oxygen bond, Fig.~\ref{fig1}a inset.
TEMPO exhibits an antiferromagnetic transition at around 2.4\,K~\cite{Matsumoto2017}, due to the 1D Heisenberg antiferromagnet exchange interaction of spins in a linear chain with nearest-neighbour interactions~\cite{Kebukawa1985}.
Its radicals are linearly arranged via intermolecular bonding within the nitrogen-oxygen bond~\cite{Yamauchi1971}. Cooling induces short-range ordering in a chain-like manner due to overlapping $\pi$-electron orbitals, followed by long-range 3D-ordering. By using all-electrical propagating spin-wave spectroscopy (AEPSWS) in a broad temperature interval, we show that coherent excitations can propagate at high magnetic fields even above the N\'eel temperature. We link the observed excitations to paramagnons hybridised with electromagnetic waves (paramagnon–polaritons).

\begin{figure*}[t!]
\centering
\includegraphics[width=1\textwidth]{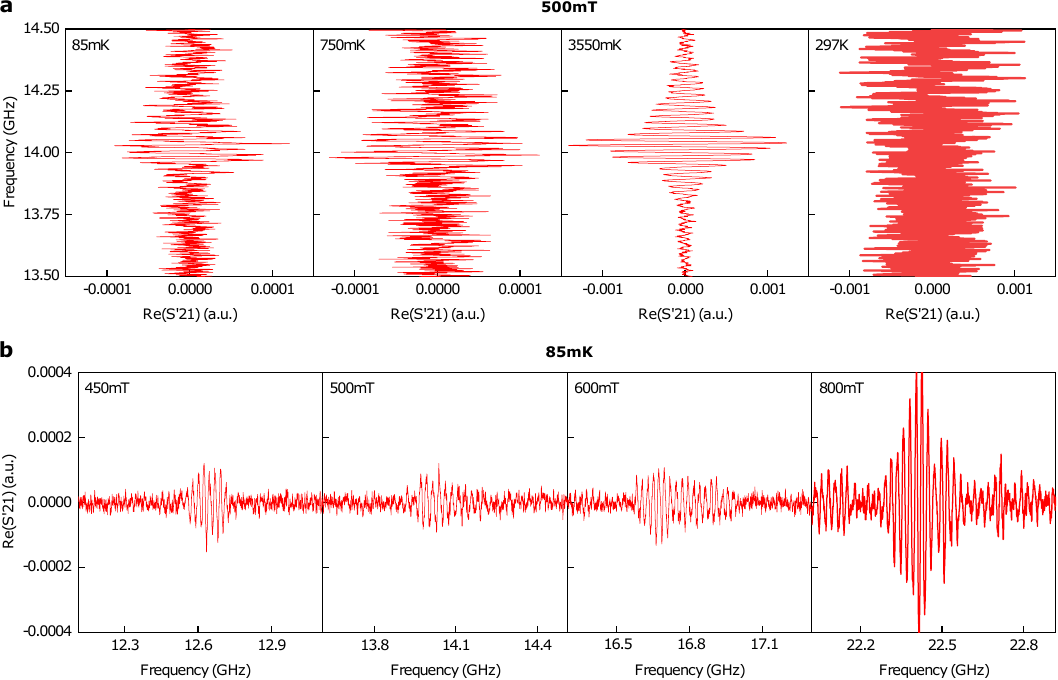}
\caption{\textbf{Temperature- and field-dependent AEPSWS of TEMPO.}~\textbf{a,}~Background corrected real part of S$'_{21}$ at different temperatures and fixed external magnetic field of $B_{\parallel}=500\mathrm{mT}$. The temperature is swept from 85 to 3550\,mK, and a reference measurement at 297\,K is taken. A coherent signal is visible for the cryogenic temperatures. The signal amplitude increases by one order of magnitude as the N\'eel point is approached (note different amplitude on abscissa), while the signal is not visible at room temperature. The SNR ratio is largest around the N\'eel point for the propagating paramagnon-polaritons, possibly due to the protection from temperature broadening near the quantum critical point (see main text).~\textbf{b,}~Background corrected real part of part of S$'_{21}$ at different external magnetic fields, measured at 85\,mK. The propagating signal shifts as the magnetic field increases. An amplitude increase is observed with increased field due to the increased effective magnetisation.
}
\label{fig3}
\end{figure*}


First, we characterised the magnetisation of bulk TEMPO, using vibrating-sample magnetometry between 2–290\,K and 0–9\,T (Fig.~\ref{fig1}b) using a resin 3D-printed sample holder (Methods; Supplementary~Section~I). With increasing field and decreasing temperature, the magnetisation approaches saturation, reaching about $24\,\mathrm{kA\,m^{-1}}$ at 2\,K and 9\,T, consistent with ref.~\cite{Karimov1968}, which also suggests that at 50\,mK saturation is expected near 6\,T. To estimate the effective magnetisation for our entire temperature range, we fit the 2\,K magnetisation data with a Brillouin function and extrapolate to 50\,mK (Fig.~\ref{fig1}c). For more details, see Methods and Supplementary~Section~II. For example at 0.5\,T this gives about $2.2\,\mathrm{kA\,m^{-1}}$. Thus, for sub-tesla fields ($\leq 1$\,T) we operate well below full saturation across the entire temperature regime.
Figure~\ref{fig1}d shows the measured magnetisation with respect to temperature for three different magnetic fields. The inset illustrates the flattening of the magnetisation curves below about 3\,K, indicating the phase transition between paramagnetic and antiferromagnetic states. The black dashed line indicates the N\'eel temperature $T_{\mathrm{N}}=2.4\,\mathrm{K}$ literature value, agreeing with our measurements.

To analyse TEMPO further and to demonstrate paramagnon–polariton propagation, we uniformly fill a glass capillary with TEMPO, seal it, and mount it on a PCB (Fig.~\ref{fig1}a). The PCBs carry either a grounded stripline for EPR or a two-port microwave antenna for propagation (4 and 8\,mm, shown in Fig.~\ref{fig1}a). The TEMPO column inside the capillary sits \mbox{$\sim 0.2$\,mm} above the conductors, set by the capillary wall thickness. Measurements are performed in a dilution refrigerator between 50\,mK and \mbox{$\sim 3.6$\,K}, enabling EPR, continuous-frequency and time-resolved pulsed AEPSWS. A separate setup is used at room temperature (Methods; Supplementary~Sections~I and III).

Figure~\ref{fig1}e shows EPR absorption spectra at $B_{\parallel}=0.5$\,T using the grounded stripline. The traces are background-corrected. A single Lorentzian fits 297\,K, while a double Lorentzian fits cryogenic spectra. 
The central resonance peaks align well with the standard formalism (see Supplementary~Section~IV) for the Zeeman splitting.
However, the linewidth broadens from $\sim 42$\,MHz (297\,K) to $\sim 67$\,MHz~(0.75\,K) and the centre frequency shifts $14.066\!\to\!14.014$\,GHz, giving Land\'e factors $g:2.0100\!\to\!2.0025$. 
At room temperature, for bulk TEMPO 2.0100 is consistent with ref.~\cite{Kobayashi2008}. At cryogenic temperatures, the g-factor is close to that of a free electron (2.0023), indicating that the electron on the oxygen atom is well shielded. We cannot resolve the full g-tensor for bulk TEMPO and assume the dominant effect arises from the glass capillary's long axis (the direction of propagation). Unlike previous reports by {\it Matsumoto and Shimosaka}~\cite{Matsumoto2017}, who observed an increase in the g-value from about 2.0075 at room temperature to 2.0095 at 10\,K, our results show a reversal at low temperatures, suggesting the electron becomes uncoupled as the temperature approaches the critical point, leading to narrow line widths and potential protection against temperature broadening~\cite{Sushkov2014, Merchant2014}.
The low temperature splitting is consistent with hyperfine coupling (Fermi contact)~\cite{Kobayashi2005}, see Supplementary~Section~IV. At 50\,mK, this feature appears as a broader double peak. Additional broadening at the lowest temperature may arise from coupling to capillary/guide modes. A detailed analysis is beyond the scope of this manuscript.

We also observe that the absorption increases by about twenty times when comparing cryogenic to room-temperature measurements (see Fig.~\ref{fig1}e), which can be attributed to the enhanced effective magnetisation at cryogenic temperatures.
These data set the loss and polarisation for the propagation measurements.

We continue to demonstrate propagating paramagnon-polaritons both above and below the N\'eel temperature.
Figure~\ref{fig1}f shows the propagation of a paramagnon-polariton at \mbox{$T=3.55$\,K}, i.e.~above \mbox{$T_{\mathrm N}=2.4$\,K}, at $B_{\parallel}=500$\,mT and an antenna spacing of 4\,mm (geometry in Fig.~\ref{fig1}a). We plot the linear magnitude (black), the real part (red), and the imaginary part (blue).
The raw $S_{21}(f)$ is corrected using a shifted reference obtained by detuning the external field so that the resonance lies outside the measurement band (typically by $\sim 70$\,mT). The fixed $\pi/2$ relation between Re$(S'_{21})$ and Im($S'_{21}$) across the passband evidences phase-stable propagation of a wave, the nature of which is uncovered below.
More information on the setup for the propagation measurements can be found in Methods and Supplementary~Section~III.


Having established propagation in the paramagnetic phase, we next map its temperature and field dependence for a parallel applied field $B_{\parallel}$ (Fig.~\ref{fig1}a). Figure~\ref{fig3}a shows the background-corrected real part of $S'_{21}(f)$ at fixed $B_{\parallel}=500$\,mT for temperatures from 85 to 3550\,mK, with a 297\,K reference. A coherent propagation band is visible up to 3.55\,K but is absent at room temperature. The amplitude increases by about an order of magnitude as it approaches the Kelvin range, and the SNR is maximised there. Although magnetisation is higher at millikelvin temperatures, EPR indicates increased absorption at mK, which would normally reduce transmission. One possible explanation for the SNR peak near Kelvin is protection against temperature broadening near the quantum critical regime/point~\cite{Sushkov2014, Merchant2014}. Further measurements will be needed to clarify this behaviour.

Figure~\ref{fig3}b shows the background-corrected real part of $S'_{21}(f)$ at 85\,mK for different $B_{\parallel}$. The propagating passband shifts with field, while the central resonance follows the Zeeman relation (Supplementary~Section~IV), confirming the magnetic nature of the propagating signal. Below $\sim 400$\,mT the signal becomes undetectable, consistent with insufficient polarisation for coherent transport. The passband amplitude increases (approximately doubles) with $B_{\parallel}$, attributable to the larger effective magnetisation.
Above the N\'eel temperature, we observe similar field dependence for both $B_{\parallel}$ and $B_{\perp}$. In this regime, transmission peak position follows the same $\omega \approx \gamma B$ relation, while the amplitude is largest at smaller fields for $B_{\parallel}$ and decreases with increasing $B_{\perp}$. Additional discussion is provided in Supplementary~Section~V.

\begin{figure}[th!]
\centering
\includegraphics[width=0.34\textwidth]{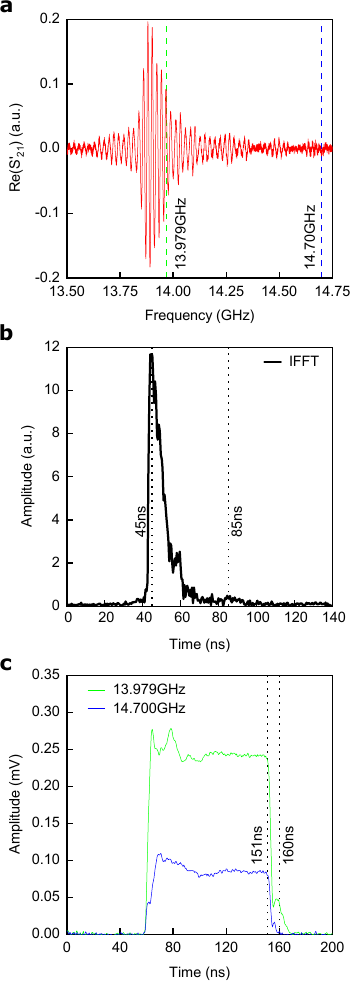}
\caption{\textbf{Time-resolved measurements.}~\textbf{a,}~Continuous-frequency measurement (real part) for $B_{\mathrm{\perp}}=500\,\mathrm{mT}$ and 55\,mK, at 4\,mm antenna spacing. \textbf{b,}~Inverse fast Fourier transformation (IFFT) of the continuous-frequency signal in~a. The first peak at 45\,ns is due to electromagnetic delay/leakage in the setup. The signal is visible until about 85\,ns, resulting in a spin-wave propagation time of about 40\,ns. \textbf{c,}~Pulsed measurement with absolute measured amplitudes for comparing resonant and off-resonant excitation of the spin wave, as shown in~a. A 100\,ns pulse at a repetition rate of 1\,ms is used. Consistent with the IFFT, a propagation time of approximately 9\,ns is observed when comparing on-resonant and off-resonant signals. From the phase periodicity in the continuous-frequency data and from time-of-flight, we estimate $v_{\mathrm g}\!\approx\!100$–$440\,\mathrm{km\,s^{-1}}$.
}\label{fig4}
\end{figure}

To confirm that the field-dependent signal arises from propagation rather than electromagnetic leakage between ports (geometry in Fig.~\ref{fig1}a), we perform time-resolved measurements and emulation (example in Fig.~\ref{fig4}). At millikelvin temperatures, the pulsed readout suppresses leakage more strongly than at kelvin temperatures, improving contrast between leakage and the travelling wave.
As a reference, we first record the continuous-frequency spectrum of the propagating band in (Fig.~\ref{fig4}a). Because time-domain measurements require higher gain, the apparent amplitude is larger than in earlier spectra (modified setup in Supplementary~Section~III). The inverse fast Fourier transform~(IFFT) of Fig.~\ref{fig4}a in Fig.~\ref{fig4}b shows an imitate peak at $\sim 45$\,ns from through-path/leakage and cryostat delay, followed by an envelope extending to a maximum at $\sim 85$\,ns, suggesting a propagation time of $\sim 40$\,ns over 4\,mm and a group velocity of $v_{\mathrm g}\!\approx\!100\,\mathrm{km\,s^{-1}}$

For pulsed measurements, we choose two frequencies in Fig.~\ref{fig4}a: on-resonant (green dashed) and off-resonant (blue dashed). With 100\,ns pulses at a 1\,ms repetition rate, the time traces (Fig.~\ref{fig4}c) show a twofold amplitude increase on resonance and an additional $\sim 9$\,ns arrival delay relative to off resonance. An overall offset arises from diode latency. Therefore, from the phase periodicity in the continuous-frequency data (Fig.~\ref{fig4}a,b), and from time-of-flight (Fig.~\ref{fig4}c), we can estimate a group velocity band of about $v_{\mathrm g}\!\approx\!100 - 440\,\mathrm{km\,s^{-1}}$ for the slowest branch.
In Supplementary~Section~V, we demonstrate that the on-resonant signal can be analytically reconstructed by convolving the continuous-frequency measurement with the analytical pulse function, confirming the consistency of IFFT-based emulation with direct pulse measurements. This analysis enables us to examine propagation velocities at few-K temperatures, where direct pulse measurements yield inconclusive results due to weak signal levels. This analysis reveals a delay time of $\sim 14$\,ns at 3550\,mK for the 4\,mm antenna spacing, which corresponds to the velocities of $\sim 285\,\mathrm{km\,s^{-1}}$. More details are discussed in Supplementary~Section~V.

The 8\,mm antenna spacing geometry corroborates these results but within a narrower passband, see Supplementary~Section~II for setup modification and~V for further results: the IFFT gives $\sim 49$\,ns and the pulsed trace $\sim 40$\,ns, on resonance, the amplitude increases by $\sim 70\times$. These values imply a group velocity band of $v_{\mathrm g}\!\approx\!160 - 200\,\mathrm{km\,s^{-1}}$, consistent with the 4\,mm device, and allowing us to narrow the bound for the actual group velocity of the propagating paramagnon-polaritons.

\begin{figure}[t!]
\centering
\includegraphics[width=0.34\textwidth]{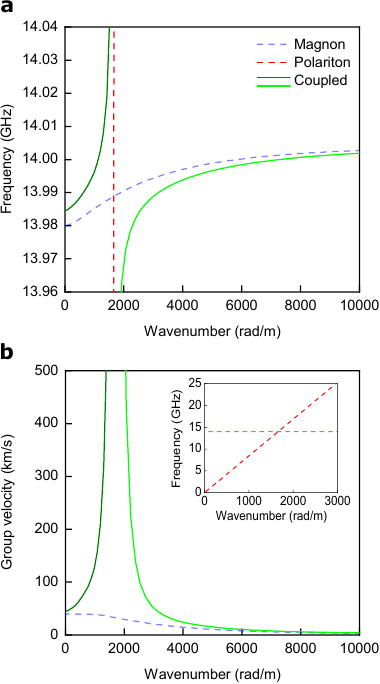}
\caption{\textbf{Theoretical dispersion calculation.}~\textbf{a,}~Close-up view of hybridised magnon-polariton dispersion curves (solid green lines); uncoupled magnon and polariton dispersions are shown in dashed (blue, red). \textbf{b,}~Corresponding group velocity of paramagnons (dashed) and coupled magnon-polaritons (solid). The inset shows a full-scale view of the dispersions. Data are for $B_\bot = 500\,$mT and $M_s = 2.2\,$kA/m. Calculation details are given in Supplementary~Section~II.
}\label{figTheory}
\end{figure}

Summarising the experimental data, we observed field-tunable, thus magnetic-origin, phase-coherent wave propagation with high velocities exceeding $100\,\mathrm{km\,s^{-1}}$. Theoretical and micromagnetic calculations of paramagnon dispersion (see details in Supplementary~Section~II) reveal that their velocities can reach just $54\,\mathrm{km\,s^{-1}}$ for $B_\parallel$ and even smaller values for $B_\bot$; results for $B_\bot$ are presented in Fig.~\ref{figTheory}. Thus, the observed transmission originates not from pure magnetic excitations, paramagnons, but from hybridised paramagnons with other excitations. The most evident candidates are magnon-polaritons.

Generally, magnon–polaritons arise from the coherent hybridisation of collective spin precession with microwave photons, producing normal-mode splitting, field-tunable dispersion, and, when dissipation and coupling are imbalanced, non-Hermitian effects at exceptional points. Experiments have observed exceptional points in cavity magnon–polaritons~\cite{Zhang2017}, demonstrated a dynamic-Hall analogue with coherently controlled logic in a planar cavity device~\cite{Rao2019}, and achieved coherent control by dynamically encircling an exceptional point~\cite{Lambert2025}. 


In Supplementary~Section~II, we elaborate further on the coupling of paramagnons to polaritons. An example of the coupled dispersion is shown in Fig.~\ref{figTheory}a. Mode anti-crossing results in rapid variation in the hybridised waves' velocity with wavenumber near the hybridisation points, which can reach hundreds of km/s (Fig.~\ref{figTheory}b). In reality, a narrow magnon band, relatively weak magnon-polariton coupling (due to weak magnetisation), which is smaller or comparable to the inverse magnon lifetime, limits accessible velocities of magnon-polaritons. Simulations of magnon-polariton pulse propagation (Supplementary~Section~II) demonstrate delays of about 15\,ns, consistent with the experimental observations. Together with the other key observation mentioned above, this leads us to conclude that the signal transmission is of a magnon-polariton nature. 

Note that the uncovered physics is valid both above and below the N\'eel temperature. Despite the phase transition at $T_{\mathrm{N}}$, the static magnetisation varies weakly, as seen in Fig.~\ref{fig1}d. In our experimental setup, only small wave-number regimes are accessible, within which exchange interactions are negligible and magnon properties are predominantly governed by dipolar interactions, analogous to the behaviour observed in the paramagnetic phase. The only difference is in the notations; above $T_{\mathrm{N}}$ the magnetic excitations are paramagnons, while below, dipolar-dominated antiferromagnetic magnons, which both couple to the polariton in the same manner via net dynamic dipolar fields.

We have demonstrated coherent, phase-stable propagation of paramagnon–polaritons in the free radical TEMPO up to 3.55\,K, over 4 and 8\,mm, with group velocities exceeding $100\,\mathrm{km\,s^{-1}}$ under sub-tesla fields. This establishes organic radicals as viable hosts for guided spin transport in the paramagnetic phase.

Building on these findings, the tunable parameters governing the passband and group velocity serve as critical control levers. Chemical modulation of spin density, hyperfine interactions, and g-factor anisotropy offers a pathway to narrow EPR linewidths and enhance magnetic filling while maintaining paramagnetic conditions. The versatility of the TEMPO family (including 4-hydroxy-TEMPO/Tempol, 4-oxo-TEMPO/Tempone, or 4-amino-TEMPO) supports systematic tuning, miniaturisation into patterned guides, and the development of 1D nanochannels, which could realise compact paramagnetic conduits for on-chip routing~\cite{Kobayashi2005} and interfaces for high-density information processing and hybrid quantum devices~\cite{Gatteschi1994}. 

In the near term, a field-tunable, phase-stable band within a sub-tesla paramagnetic regime enables reconfigurable delay lines, interferometric routing, and low-loss cryogenic interconnects compatible with superconducting microwave hardware. Looking further ahead, integrating this guided transport with organic electronic platforms, leveraging flexible substrates and solution processing, could extend magnonic functionalities beyond inorganic magnets, while preserving chemical control over dispersion and dissipative losses.



\newpage
\printbibliography[heading=subbibliography, title={References}, segment=\therefsegment]
\end{refsegment}

\begin{refsegment}
\section*{Online Methods}\label{sec11}
\noindent\textbf{Sample preparation}\\
Commertially avaiblable 2,2,6,6-Tetramethylpiperidin-1-oxyl (molar weight 156.25\,g/mol) with a purity of $\geq97.5\,\%$ is molten to a constant temperature of $65^{\circ}C$. Capillary forces are then used to fill a capillary lab tube (inner diameter 1.1\,mm, outer diameter 1.5\,mm) of about 15\,mm in length. The ends are then sealed. For the continuous-frequency and pulsed measurements, the sample is subsequently mounted on a PCB with grounded microwave antennas, with a width of 25\,$\mu$m and a spacing of 4\,mm, and 8\,mm for the pulsed measurements. The PCB and cryogenic setup allow operation up to 40\,GHz. More details on the sample preparation can be found in Supplementary~Section~I. 

\vspace{0.3cm}
\noindent\textbf{EPR measurements}\\
For the EPR measurements, the glass capillary  sample is affixed to a stripline high-frequency PCB. The stripline PCB consists of a 200\,$\mu$m coplanar waveguide (dielectric constant 3.26) with 150\,$\mu$m spacing on either side from the ground line and is matched to 50\,$\Omega$, see Supplementary Section I and IV.

\vspace{0.3cm}
\noindent\textbf{Vibrating sample magnetometer measurements}\\
The magnetisation measurements are performed in a vibrating sample magnetometer~(VSM) attachment within a physical property measurement system ({\it Quantum Design}). The VSM allow a continuous magnetisation measurement within the temperature range of 2-300\,K in a homogenous magnetic field of 0-9\,T. A cylindrical 3d-printed resin sample holder, filled with TEMPO, was built only for this specific measurement. More details on this sample holder can be found in Supplementary~Section~I.

\vspace{0.3cm}
\noindent\textbf{Cryogenic propagating spin wave measurements}\\
The continuous-frequency and pulsed propagating paramagnon-polariton measurements are performed in a cryogenic-free dilution refrigerator system ({\it Bluefors-LD250}), which reaches base temperatures below 10\,mK at the mixing chamber stage. The sample space possesses a base temperature of about 16\,mK. During operation, the sample space heats up to about 50-85\,mK. The dilution refrigerator can be operated continuously, depending on the used magnetic field, between the base temperature and 3.6\,K.
The input signal is transmitted and collected from the sample (ports 1 and 2 in Fig.~\ref{fig1}a main text), using high-frequency copper and superconducting wiring for the continuous-frequency propagating spin-wave spectroscopy. The signals are collected with a 70 GHz vector network analyser ({\it Anritsu~VectorStar~MS4647B}). For time-resolved pulsed measurements, the dilution refrigerator setup is modified by adding a pulsed microwave source {\it Anritsu~MG3692C}, fast detection diode {\it OMNIspecta~Mod20760}, RF-bandpass filter {\it Mini-Circuits~VHF-6010+}, and oscilloscope {\it Teledyne~HDO6034}. More details on the cryogenic setup can be found in Supplementary~Section~III.

Note, in this work, the spins are aligned by an external magnetic field in TEMPO, resulting in a positive net magnetisation that persists above $T_{\mathrm{N}}$. Thus, we refer to these propagating magnons as paramagnon-polaritons, characterised by an additional polaritonic contribution resulting from their coupling to the electromagnetic wave of the signal. More details are provided in the main text.

The reference room temperature measurements are conducted using a home-built setup. This setup comprises a vector network analyser ({\it Rohde \& Schwarz~ZNA67}) connected to an H-frame electromagnet {\it GMW 3473-70}, which includes an 8 cm air gap for various measurement configurations and magnet poles with a diameter of 15\,cm to produce a sufficiently uniform biasing magnetic field. The electromagnet is powered by a bipolar power supply {\it ICO BPS-85-70E}C, generating up to 0.9\,T at the 8\,cm air gap. The input powers are adjusted to achieve identical power levels at the sample as those in the cryogenic measurements, accounting for cable losses and the previously mentioned attenuators while excluding impedance mismatches that may arise from temperature changes. 

The fixed $\pi/2$ relation between Re$(S'_{21})$ and Im($S'_{21}$) across the passband evidences phase-stable propagation of paramagnon–polaritons in our sample~\cite{Vlaminck2010, Vanatka2021, Knauer2023}.

\vspace{0.3cm}
\noindent\textbf{Brillouin fit}\\
For our temperature-dependent Brillouin fit~\cite{Barak1992, Serha2024} (see Fig.~1c main text, red) of the 2\,K data, we use the experimentally obtained effective saturation magnetisation of $ M_0=24 \mathrm{kA/m}$ (see Fig.~1b main text) and keep the molecular field coefficient as a free fitting parameter, which gives us $ \lambda \approx -183$. More data and information can be found in Supplementary~Section~II.

\vspace{0.5cm}
\printbibliography[heading=subbibliography, title={References online methods}, segment=\therefsegment]
\end{refsegment}

\section*{Data availability}
The data presented in the current study are available from the corresponding authors upon reasonable request.

\section*{Acknowledgements}
This project has received funding from the European Union’s Horizon 2020 research and innovation programme under the Marie Sklodowska-Curie grant agreement No.~101025758, project OMNI. 
This research was funded in part by the Austrian Science Fund (FWF) [10.55776/I6568], project Paramagnonics.
The IMAG team acknowledges the support by the NAS of Ukraine project No.~0123U104827. 
This work at M{\"u}nster University was supported by
the Deutsche Forschungsgemeinschaft (DFG, German Research Foundation) – project number 523303368.

\vspace{0.5cm}
\noindent\textit{Corresponding authors:}\\
S.~Knauer: knauer.seb@gmail.com\\
A.~V.~Chumak: andrii.chumak@univie.ac.at\\

\section*{Author contributions}
S.K.~conceived the idea of TEMPO as material, prepared the sample, planned and performed the experiments, and analysed the data. S.K.~and D.Sch.~build the dilution refrigerator setup. R.O.S.~performed the VSM measurements and supported the experiment and data analysis. D.S.~and R.V.~provided theoretical support and performed the dispersion and pulse propagation calculations. A.N.~provided theoretical support for the magnetisation measurements. S.O.D.~and A.V.C.~conceived the propagating paramagnon idea and led the paramagnonic project. S.K.~wrote the first version of the manuscript. All authors discussed the results and contributed to the writing of the manuscript.

\section*{Competing interests}
The authors declare no competing interests.

\end{document}